\documentclass[%
reprint,
amssymb,
aps,
prl,
floatfix,
]{revtex4-2}

\usepackage[utf8]{inputenc}
\usepackage{graphicx}
\usepackage{dcolumn}
\usepackage{array}
\usepackage{amssymb}
\usepackage{amsmath}
\usepackage{multirow}
\usepackage{braket}
\usepackage[toc,page]{appendix}
\usepackage{bm}
\usepackage[normalem]{ulem}
\renewcommand{\vec}[1]{\bm{#1}}
\usepackage[hang]{footmisc}
\usepackage{tikz}
\usepackage{color}
\usepackage{hyperref}
\hypersetup{colorlinks, 
	linkcolor={blue!75!black!80!yellow},
	citecolor={blue!75!black!80!yellow}, 
	urlcolor={blue!75!black!80!yellow}
	}
\usepackage{autonum}
\makeatletter \renewcommand\@make@capt@title[2]{%
\@ifx@empty\float@link{\@firstofone}{\expandafter\href\expandafter{\float@link}}%
\sffamily{\textbf{#1}}\@caption@fignum@sep#2 }

\begin{document}
\title{Ab initio linear-response approach to vibro-polaritons in the cavity Born-Oppenheimer approximation}

\author{John Bonini, and Johannes Flick}
\affiliation{Center for Computational Quantum Physics, Flatiron Institute, 162 5th Ave., New York, 10010  NY,  USA}
\date{\today}
\begin{abstract}
  Recent years have seen significant developments in the study of strong light-matter coupling including the control of chemical reactions by altering the vibrational normal modes of
  molecules. In the vibrational strong coupling regime the normal modes of the
  system become hybrid modes which mix nuclear, electronic, and photonic
  degrees of freedom. First principles methods capable of treating light and
  matter degrees of freedom on the same level of theory are an important tool in
  understanding such systems.
  In this work, we develop and apply a generalized force constant matrix approach to the
  study of mixed vibration-photon (vibro-polariton) states of molecules based on the cavity Born-Oppenheimer approximation and quantum-electrodynamical density-functional theory. With this
  method vibro-polariton modes and infrared spectra can be computed
  via linear response techniques analogous to those widely used for conventional vibrations and phonons.
  We also develop an accurate model that highlights the consistent treatment of cavity coupled electrons in the vibrational strong coupling regime.
  These electronic effects appear as new terms previously disregarded by simpler models.
  This effective model also allows for an accurate extrapolation of single and two molecule
  calculations to the collective strong coupling limit of hundreds of molecules.
  We benchmark these approaches for single and many CO$_2$ molecules coupled to a
  single photon mode and the iron-pentacarbonyl Fe(CO)$_5$ molecule coupled to a few
  photon modes. Our results are the first ab-initio results for collective vibrational
  strong coupling effects. This framework for
  efficient computations of vibro-polaritons paves the way to a systematic
  description and improved understanding of the behavior of chemical systems in
  vibrational strong coupling.
\end{abstract}
\maketitle

\section{Introduction}

Recent experimental progress in the field of polaritonic chemistry has
demonstrated the possibilities of altering chemical and material properties with
the strong coupling of electromagnetic fields and vibrational degrees of
freedom. 
In this vibrational strong coupling regime, light and matter degrees of freedom hybridize forming vibro-polaritons~\cite{ebbesen2016}.
It has been demonstrated that in this regime coupled cavity photons can be tuned to influence
chemical reactivity~\cite{ebbesenTilting}, vibrational energy
redistribution~\cite{Xiang2020}, optical spectra~\cite{George2016,kadyan2021},
Raman spectra~\cite{Shalabney2015}, two-dimensional
spectroscopy~\cite{Xiang4845}, relaxation dynamics~\cite{Grafton_2021},
ultrafast thermal modification~\cite{Liu2021}, and even
superconductivity~\cite{thomas2019exploring}, among others. These experimental works have been complemented by various theoretical efforts~\cite{flick2017,martinez2018,galego2019cavity,li2020resonance,li2020,Campos2020,li2021collective,vibok2021}, one development in particular to
describe these experiments is the introduction of effective vibro-polariton
Hamiltonians~\cite{George2016,kadyan2021,fischer21_groun_state_proper_infrar_spect,hernandez2019},
that include the vibrational degree of freedom via normal modes (vibrations in
molecular or phonons in solid-state systems). These normal modes can be obtained
e.g. experimentally from infrared spectroscopy~\cite{George2016,kadyan2021}, or
numerically from first principles using electronic-structure theory
methods~\cite{gonze1997, Baroni_2001}. Although conventional electronic structure
methods are not directly applicable to the light-matter strong coupling regime
due to their negligence of the quantum electromagnetic field, here they can be
used to calculate the vibrational normal modes of the matter system based on the force
constant matrix. These vibrational normal modes are then coupled to the photon modes of the electromagnetic
field. Such Hamiltonians have been applied successfully to describe various
experimental findings~\cite{George2016,kadyan2021,hernandez2019}. One limitation
of these vibro-polariton Hamiltonians that only include vibrational modes and
photon modes explicitly is that self-consistent effects of the electron-photon interaction 
are neglected. In addition, this description usually aims at including only the
relevant degrees of freedom of the system explicitly. While for simpler systems
the relevant degrees of freedom can be known beforehand, in general and for more
complex situations these variables are not always known.

An alternative route to simulate vibrational strong coupling
is offered by first principles methods that treat the full
matter-photon Hamiltonian explicitly. Examples include the generalization of
Hartree-Fock~\cite{Rivera2019,haugland2020coupled}, QED coupled-cluster (QED-CC)
theory~\cite{mordovina2020polaritonic,haugland2020coupled,pavosevic2021polaritonic}, and
quantum-electrodynamical density-functional theory
(QEDFT)~\cite{tokatly13_time_depen_densit_funct_theor,
  ruggenthaler14_quant_elect_densit_funct_theor}. In the QEDFT framework,
vibrational strong coupling has been simulated in the time-domain capturing the
dynamics of the system to analyze optical
spectra~\cite{flick18_cavit_correl_elect_nuclear_dynam}, or chemical
reactivity~\cite{schafer2021shining}, but the full framework to describe
vibrational strong light-matter coupling within linear-response theory has not yet been developed. While
explicit calculations in the time-domain have their advantages for simulating
complex and anharmonic dynamics, information about vibro-polaritonic modes can be obtained from
linear-response calculations more efficiently. One limitation of these
first principle methods is their relatively high computational cost, which
effectively limits calculations to the single or few molecule limit, which is
the opposite limit of experiments in the collective strong coupling regime.

In this work, we introduce an efficient framework to calculate properties of
systems under vibrational strong coupling from first principles. We introduce
the generalized force constant matrix, where eigenvectors and eigenvalues give
rise to vibro-polaritonic normal modes of the correlated matter-photon system and the
frequencies of the vibro-polaritons. In addition, we develop an accurate
effective model that includes light-matter feedback terms that have been
previously disregarded. We show that this effective model allows for
extrapolation of first principle calculations to the collective strong coupling
regime. We exemplify these methods by calculating optical spectra for single and
many CO$_2$ molecules in optical cavities, as well as for the iron-pentacarbonyl
Fe(CO)$_5$ coupled to a multi-photon mode setup.

\begin{figure}
  \label{fig:schematic}
  \includegraphics[width=\linewidth]{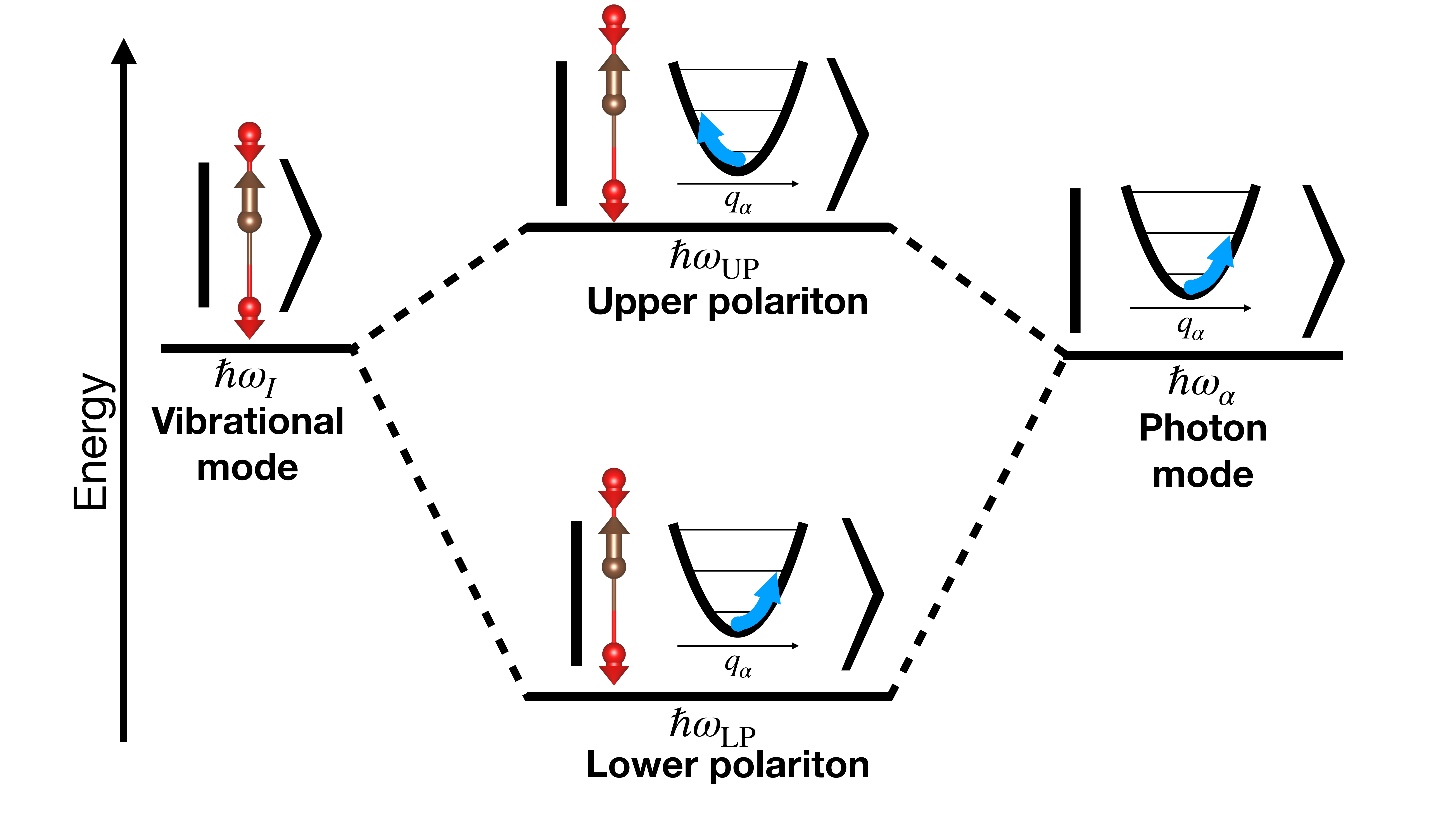}
  \caption{
  Schematic representation of vibro-polaritonic excitations in an optical cavity.
  On the left an infrared active vibrational excitation of CO$_2$ is depicted at a particular energy level. 
  On the right a particular photon mode of the cavity is depicted as an excitation of photon displacement coordinate $q_\alpha$ within a harmonic potential.
  Under strong coupling these vibration and photon modes hybridize leading to upper and lower polaritons as depicted by the two states in the center. 
  Note that the eigenvectors of these hybrid states have opposite signed $q_\alpha$ components as depicted with the blue arrows.}
\end{figure}

\section{Theory of vibro-polaritons}
\label{sec:theory}
In the following section, we develop the framework to describe vibro-polaritons in the linear-response regime from first principles. We start by discussing the Hamiltonian for light-matter coupled systems in the
length gauge and in the dipole
approximation~\cite{faisal1987,flick17_cavit_born_oppen_approx_correl,flick18_cavit_correl_elect_nuclear_dynam}.
For the vibrational strong coupling regime, it has been shown that the cavity
Born-Oppenheimer approximation (CBOA) can yield an accurate description of the
system~\cite{flick2017,flick17_cavit_born_oppen_approx_correl,galego2019cavity,li2020resonance,Campos2020}.
This method is
based on the adiabatic approximation that allows separation of the electronic
degrees of freedom from the nuclear and photonic degrees of freedom.
As a
consequence, the photonic degrees of freedom are described as analogous to the
nuclear degrees of freedom in the conventional Born-Oppenheimer approximation~\cite{BornHuang}. With
this framework the nuclear-photon dynamics of a set of $N_\text{nuc}$ nuclei
with coordinates
$\vec{\underline{R}} = \left(R_{1x}, R_{1y}, R_{1z}, R_{2x} ... R_{I\kappa} ... \right)$
and $\mathcal{N}_\text{pt}$ photon modes with photon displacement coordinates
$\underline{q} = \left(q_1, q_2, ... q_{\alpha} ... \right)$ is given by
the following Hamiltonian
\begin{align}
\label{eqn:cboa-ham-npt}
    \left(\hat{T}_\text{nuc} + \hat{T}_\text{pt} + E_i(\vec{\underline{R}}, \underline{q})\right)\Phi_j(\vec{\underline{R}}, \underline{q}) = \epsilon_j \Phi_j(\vec{\underline{R}}, \underline{q})
\end{align}
with nuclear and photonic kinetic energies $\hat{T}_\text{nuc}$ and
$\hat{T}_\text{pt}$, respectively and $E_i$ denotes the cavity Born-Oppenheimer
(CBO) potential-energy surface for the $i$th electronic energy level of the system. In practice, we can obtain the CBO
potential-energy surfaces from diagonalizing the electronic Hamiltonian of $N_e$
electrons that now parametrically depends on the nuclear and the photonic
coordinates with
\begin{align}
    \label{eqn:e-schrodinger}
     \hat{H} (\vec{\underline{R}}, \underline{q}) \Psi_i(\vec{\underline{r}},\vec{\underline{R}}, \underline{q}) = E_i(\vec{\underline{R}}, \underline{q}) \Psi_i(\vec{\underline{r}},\vec{\underline{R}}, \underline{q})
\end{align}
where 
\begin{equation}
\label{eqn:cboa-ham}
  \begin{split}
  \hat{H} (\vec{\underline{R}}, \underline{q})&= \hat{T}_e
  + \hat{V}_{e-e}(\vec{\underline{r}})
  + \hat{V}_{e-\mathrm{nuc}}(\vec{\underline{r}}, \vec{\underline{R}})
  + \hat{V}_{\mathrm{nuc} - \mathrm{nuc}}(\vec{\underline{R}})\\
  &+ \hat{V}_{\mathrm{pt} - \vec{\mu}}(\vec{\underline{r}},\vec{\underline{R}}, \underline{q})\,.
  \end{split}
\end{equation}
Here, $\hat{T}_e$ describes the electronic kinetic energy, $ \hat{V}_{e-e}$ the
electron-electron interaction, $ \hat{V}_{e-\mathrm{nuc}}$ the electron-nuclear
interaction, and $\hat{V}_{\mathrm{nuc} - \mathrm{nuc}}$ the nuclear-nuclear
interactions, respectively. In Eq.~\ref{eqn:cboa-ham}, we include the matter-photon coupling by
\begin{equation}
  \label{eqn:coupling}
  \hat{V}_{\mathrm{pt}-\vec{\mu}}(\vec{\underline{r}},\vec{\underline{R}}, \underline{q}) =
  \frac{1}{2}\sum_{\alpha=1}\omega_{\alpha}^{2}(\hat{q}_{\alpha}
  - \frac{\vec{\lambda}_{\alpha}}{\omega_{\alpha}}\cdot\hat{\vec{\mu}})^{2}
\end{equation}
where the $\alpha$ runs over photon modes, the photon displacement coordinate $q_\alpha$ couples to the electronic
and nuclear dipole moment operator, which is given by
$\hat{\vec{\mu}} = \sum_{I}eZ_{I}\vec{R}_{I} - e \sum_{i}\hat{\vec{r}}_{i}$,
where $\hat{\vec{r}}$ is the electronic position operator, $e$ describes the elementary charge and $Z_{I}$ the charge of the $I$th
nuclei. The frequency $\omega_\alpha$, and the coupling strength
$\boldsymbol\lambda_\alpha$ define the parameters of the individual photon
modes. In this work we treat the $\hat{\vec{\mu}}^2$ term in the electronic potential using a mean field approximation as described in Appendix~\ref{appendix:vmf}.

Having setup the Hamiltonian of the matter-photon system, we can proceed to
determine the vibro-polaritonic normal modes. In the first step, we define the
effective nuclear and photonic forces and calculate the equilibrium
configuration of the system.
To derive the forces acting on nuclear and photonic degrees of freedom in the presence of matter-photon coupling we apply the Hellman-Feynman theorem. The forces on the nuclei $I$ along the $\kappa$ direction, are given by
\begin{equation}
  \label{eqn:F_R}
  \begin{split}
    F_{I\kappa} = -\frac{\partial E(\vec{\underline{R}},\underline{q} )}{\partial R_{I \kappa} } &=
    \braket{\frac{d}{dR_{I \kappa}}
    (\hat{H}-\hat{V}_{\mathrm{pt}-\vec{\mu}})} \\ 
    &+ e Z_{I}\sum_{\alpha=1}\vec{\lambda}_{\alpha}\biggl(\omega_{\alpha}{q}_{\alpha}  -\vec{\lambda}_{\alpha}\cdot  \braket{\hat{\vec{\mu}}}\biggl)
  \end{split}
\end{equation}
where $E$ is the ground-state CBO energy of the system governed by the
Hamiltonian in Eq.~\ref{eqn:cboa-ham}, $R_{I\kappa}$ indicates the $\kappa$
direction component of the position of nuclei $I$, and $\braket{..}$ indicates
an expectation value evaluated using the electronic states at particular values of $\vec{\uline{R}}$ and
$\uline{q}$. There is also an effective force on the photon displacement
coordinate, which is given by
\begin{equation}
  \label{eqn:F_q}
  F_{q_{\alpha}} = -\frac{\partial E(\vec{\underline{R}},\underline{q})}{\partial q_\alpha } = - \omega_{\alpha}^{2} {q}_{\alpha}  + \omega_{\alpha} \boldsymbol\lambda_{\alpha}\cdot  \braket{\hat{\vec{\mu}}}\,.
\end{equation}
The equilibrium position with the ground state energy $E_0$ with respect to $\vec{\underline R}$ and
$\underline{q}$ is now defined by minimization of energy, as defined by Eq.~\ref{eqn:e-schrodinger}, and thus vanishing forces, i.e.
$F_{R_{I\kappa}}=F_{q_\alpha}=0$ with the electronic Hamiltonian in
Eq.~\ref{eqn:cboa-ham}.

The CBO energy of the coupled light-matter system with small perturbations
around the equilibrium configuration can be expressed as
\begin{align} 
    E(\vec {\underline{R}}, \underline{q}) &= E_0 +
    \sum_{I\kappa,J\kappa'}  \frac{1}{2}
    C^{(RR)}_{I\kappa,J\kappa'}
    \Delta R_{I\kappa} \Delta R_{J\kappa'} \nonumber\\
    &+ \sum_{\alpha,\alpha'}  \frac{1}{2}
    C^{(qq)}_{\alpha,\alpha'}
    \Delta q_\alpha \Delta q_{\alpha'} \nonumber\\
    &+ \sum_{\alpha,I\kappa} 
    C^{(qR)}_{\alpha,I\kappa}
      \Delta R_{I\kappa} \Delta q_\alpha\nonumber\\
      &+ \mathcal{O}(\Delta R_{I\kappa}^3, \Delta q_\alpha^3, \Delta q_\alpha\Delta R_{I\kappa}^2, \Delta R_{I\kappa}q_\alpha^2)   \label{eqn:E_Rq}
\end{align}
where $\Delta R_{I\kappa}$ are displacements of atom $I$ along direction $\kappa$,
$\Delta q_\alpha$ are perturbations of photon displacement $q_\alpha$, $E_0$ is the energy of the equilibrium configuration, and we
have defined the matrices
\begin{equation}
  \label{eqn:CRR}
    C^{(RR)}_{I\kappa,J\kappa'} =
    \frac{\partial^2E(\vec {\underline{R}}, \underline{q})}{\partial R_{I\kappa} \partial R_{J\kappa'}}
     = -\frac{\partial F_{R_{I\kappa}}}{\partial R_{J\kappa'}}
\end{equation}
\begin{equation}
  \label{eqn:Cqq}
    C^{(qq)}_{\alpha,\alpha'} = 
    \frac{\partial^2E(\vec {\underline{R}}, \underline{q})}{\partial q_\alpha \partial q_{\alpha'}}
     = -\frac{\partial F_{q_{\alpha}}}{\partial q_{\alpha'}}
\end{equation}
\begin{equation}
  \label{eqn:CqR}
    C^{(qR)}_{\alpha,I\kappa} = 
    \frac{\partial^2E(\vec {\underline{R}}, \underline{q})}{ \partial q_\alpha \partial  R_{I\kappa}}
     = -\frac{\partial F_{q_{\alpha}}}{\partial R_{I\kappa}}\,.
\end{equation}
The vibro-polariton eigendisplacements $\eta_m$ of the light-matter coupled system
and the vibro-polariton eigenfrequencies $\omega_{m}$ can be obtained by solving
the generalized eigenvalue problem
\begin{equation}
  \label{eqn:gen_ev_mat}
 \begin{pmatrix}
   C^{(RR)} & {(C^{(qR)})}^{T} \\
   C^{(qR)}  & C^{(qq)}
 \end{pmatrix}
 \begin{pmatrix}
   \eta_{m}^{(\uline{R})} \\
   \eta_{m}^{(\uline{q})}
 \end{pmatrix}
 =
 \begin{pmatrix}
M & 0\\
0 & \mathcal{I}
 \end{pmatrix}
 \omega_{m}^{2}
 \begin{pmatrix}
   \eta_{m}^{(\uline{R})} \\
   \eta_{m}^{(\uline{q})}
 \end{pmatrix}
\end{equation}
where $M_{I\kappa,J\kappa'} = M_{I}\delta_{IJ}\delta_{\kappa\kappa'}$, $M_{I}$ is the mass of nuclei $I$, and $\mathcal{I}$ is a $\mathcal{N}_{\mathrm{pt}}\times \mathcal{N}_{\mathrm{pt}}$ identity matrix.
\footnote{ In an effort to treat light and matter degrees of freedom on equal
  footing in the notation in this definition, we have implicitly treated
  $C^{(RR)}_{I\kappa, J\kappa'}$ as a
  $3N_{\mathrm{nuc}}\times 3N_{\mathrm{nuc}}$ matrix with only two indices so
  that
  $C^{(RR)}_{I\kappa, J\kappa'} \rightarrow C^{(RR)}_{3I+\kappa,3J+\kappa'}$
  with indexing starting at zero. Similarly we treat $C^{(qR)}$ as the
  $N_{\mathrm{photon}}\times 3N_{\mathrm{nuc}}$ matrix
  $C^{(qR)}_{\alpha,I\kappa} \rightarrow C^{(qR)}_{\alpha, 3I+\kappa}$ and $M$
  as the $3N_{\mathrm{nuc}}\times 3N_{\mathrm{nuc}}$ matrix $M_{I\kappa, J\kappa'} \rightarrow M_{3I+\kappa, 3J+\kappa'}$. }
The matrices $C$ and $\tilde{M}$ as well as generalized
eigendisplacements $\eta_{m}$ can be used to rewrite Eq.~\ref{eqn:gen_ev_mat} in a more
compact form
\begin{equation}
  \label{eqn::gen_ev_mat_compact}
  C \eta_{m} = \tilde{M} \omega_{m}^{2}\eta_{m}\,.
\end{equation}
Where now $C$ acts as a generalized force constant matrix which
includes both nuclear and photon degrees of freedom. The analogous generalized
dynamical matrix can then be defined as
\begin{equation}
  D_{ij} = C_{ij} / (\tilde{M}_{ii}\tilde{M}_{jj})^{1/2}
\end{equation}
with eigenvalues $\omega_m^{2}$ and vibro-polariton eigenvectors $U_{m}$. For a normalized set of
$U_{m}$ the eigendisplacements are related by
$\eta_{m,i}=U_{m,i}/{(\tilde{M}_{ii})}^{1/2}$, where the eigendisplacements are
normalized to obey $\eta_{m}^{T}\tilde{M}\eta_{m}=1$.

Analyzing the structure of the force constant matrix, we find a $2\times2$ block
structure of $C$ (left side of Eq.~\ref{eqn:gen_ev_mat}) reminiscent of
the electron-photon linear-response polaritonic Casida
equation~\cite{flick2019lr}. We find the matter block $C^{(RR)}$
and photon block $C^{(qq)}$ on the diagonal coupled by an off-diagonal
block $C^{(qR)}$, which introduces the matter-photon coupling. For the
case of $\lambda=0$, the off-diagonal blocks vanishes and the matter block reduces
the standard force constant matrix~\cite{Baroni_2001}. We further note that
while in the polaritonic Casida equation the photon block is strictly diagonal,
since there is no explicit photon-photon interaction present, the same is not
true for the generalized force constant matrix here. In this case, the photon
block $C^{(qq)}$ is not diagonal due to an effective photon-photon interaction between individual photon modes that originates from the electron-photon
interaction. The manifestations of this effective photon-photon interaction will
be discussed in Sec.~\ref{ssec:FeCO5}. A schematic representation for a single vibrational mode coupled to a single photon mode is given in Fig.~\ref{fig:schematic}.

We can now obtain the infrared spectrum from the eigenvectors and eigenfrequencies of the generalized dynamical matrix and the vibro-polariton mode effective charges. Both quantities are defined analogously to the case of conventional linear response theory of vibrations/phonons \cite{gonze1997}. The mode effective charge of vibro-polariton normal mode $m$ along direction $\kappa$ is given by
\begin{align}
\label{eqn:born-charge}
  Z_{m,\kappa}^* = \sum_{I}\sum_{\kappa'}\frac{\partial \braket{\hat \mu_{\kappa}}}{\partial R_I} \eta_{m, I\kappa'}^{(\uline{R})}
  + \sum_{\alpha}\frac{\partial \braket{\hat \mu_{\kappa}}}{\partial q_{\alpha}}\eta_{m, \alpha}^{(\uline{q})}\,.
\end{align}
Using these effective charges, the corresponding infrared spectrum $I$ can be constructed as
\begin{align}
    I(\Omega) = \sum_m |\vec Z_m^{*}|^2 L(\Omega, \omega_m,\delta),
\end{align}
where the peaks at the frequencies $\omega_m$ with amplitudes $Z_m$ are broadened by the Lorentzian $L(\Omega, \omega_m,\delta)$~\cite{wang2021}.

\section{Model for vibro-polaritons}

To elucidate the various microscopic contributions to the results of the full first principles theory we now develop an equivalent vibro-polaritonic model.
We first rewrite Eq.~\ref{eqn:E_Rq} with the
matter degrees of freedom rotated into a basis of uncoupled vibrational normal modes.
The nuclear displacements from the equilibrium configuration can be expressed in terms of vibrational mode amplitudes $N_{I}$ which specify the change of ionic positions.
For a general ionic displacement given by a set of $\Delta R_{I\kappa}$ the corresponding set of $N_I$ are given by
$N_I = \sum_{J\kappa} (\eta_{m, J\kappa}^{(\uline{R})}\bigl|_{\lambda=0})^\mathrm{T} \Delta R_{J\kappa}$
where $\eta^{({\uline{R}})}_{m,J\kappa}\bigl|_{\lambda=0}$ are the normal
vibrational mode eigendisplacements of the uncoupled problem at $\lambda =0$. Force constant matrix elements
can be written in terms of $N_I$ and $q_{\alpha}$ by expanding the
expectation values present in Eqs.~\ref{eqn:F_R} and \ref{eqn:F_q} to linear
order. Then the dipole expectation value reads
\begin{equation}
\braket{\hat{\vec{\mu}}} \approx \sum_{I}\frac{\partial\braket{\hat{\vec{\mu}}}}{\partial N_{I}}N_{I} + \sum_{\alpha}\frac{\partial\braket{\hat{\vec{\mu}}}}{\partial q_{\alpha}}q_{\alpha}\,.
\end{equation}
We express the first force contribution on the left side of Eq.~\ref{eqn:F_R} in terms mixed second derivatives 
given by matrices $\Theta$ and $\Xi$, defined explicitly in Appendix~\ref{appendix:model_details}.
With the above expansions and change of basis we can define the following harmonic model:
\begin{equation}
  \label{eqn:general_model}
\begin{split}
  \hat H_M &= \hat T +
  \sum_{I, J=1}\frac{1}{2}\biggl[\omega_{I}^{2}\delta_{IJ} + \Xi_{IJ}^{(\uline{\lambda})}\\
&  +\sum_{\alpha=1}\biggl(\boldsymbol\lambda_{\alpha}\cdot e\vec{Z}_{I}\biggl)\left(\boldsymbol\lambda_{\alpha}\cdot\frac{\partial\braket{\hat{\boldsymbol\mu}}}{\partial N_{J}}\right)
  \biggl]\hat N_{I} \hat N_{J}\\
&+ \sum_{\alpha, \alpha'=1} \frac{1}{2}\left(\omega_{\alpha}^{2}\delta_{\alpha\alpha'} - \omega_{\alpha}\boldsymbol\lambda_{\alpha}\cdot\frac{\partial\braket{\hat{{\boldsymbol\mu}}}}{\partial q_{\alpha'}}\right)\hat q_{\alpha}\hat q_{\alpha'}\\
&- \sum_{\alpha,I=1}\omega_{\alpha}\boldsymbol\lambda_{\alpha}\cdot\frac{\partial\braket{\hat{\boldsymbol\mu}}}{\partial N_{I}}\hat N_{I} \hat q_{\alpha}
\end{split}
\end{equation}
where $T$ includes the kinetic energy of the vibrational modes ($N_I$) and the
photon modes ($q_\alpha$). $Z_{I \kappa}$ is the ionic contribution to the uncoupled vibrational mode effective charge of Eq.~\ref{eqn:born-charge} and given explicitly by
$Z_{I \kappa}= e\sum_{J} Z_{J} \eta_{I, J\kappa}^{(\uline{R})}\bigl|_{\lambda=0}$.
Additional details on the derivation of the model can be found in
Appendix.~\ref{appendix:model_details}. 

We find that the matter-photon coupling
strength, i.e. the term proportional to $N_I q_\alpha$, depends on the
quantity $\frac{\partial\braket{\hat{\boldsymbol\mu}}}{\partial N_{I}}$. As
discussed before, the dipole moment of the system consists of two contributions,
a nuclear one and the electronic one. As a consequence the term
$\frac{\partial\braket{\hat{\boldsymbol\mu}}}{\partial N_{I}}$ also includes two
contributions: The nuclear dipole moment, as well as the change of the electric
dipole moment due to a change in nuclear configuration. Analogously, the term
$\frac{\partial\braket{\hat{\boldsymbol\mu}}}{\partial q_{\alpha}}$ describes the
change of the electric dipole moment due to a change in photon coordinate
$q_\alpha$. We note that while photon modes are not explicitly
coupled in Eq.~\ref{eqn:cboa-ham}, i.e. there is no photon-photon coupling term, the 
$\frac{\partial\braket{\hat{\boldsymbol\mu}}}{\partial q_{\alpha}}$ term 
introduces effective
photon-photon coupling in the vibro-polariton model.
Since the model
describes a set of interacting quantum harmonic oscillators, it can also be solved
analytically~\cite{Burrows2003}.

For a detailed illustration, we now consider the model of
Eq.~\ref{eqn:general_model} for a single photon mode coupled to a single vibration
mode with the relevant vibration only influencing the dipole moment along the direction of photon polarization. With these simplifications we can drop the mode indices, label the vibration mode frequency with subscript $N$ and the photon mode with subscript $q$, and treat the dipole moment $\mu$ and coupling strength vector $\lambda$ as scalars.
Then Eq.~\ref{eqn:general_model} reduces to
\begin{equation}
  \label{eqn:two_mode_model}
 \hat H_{SM} = \hat{T} + \frac{1}{2} \tilde{\omega}_N^{2}\hat N^{2}+ \frac{1}{2} \tilde{\omega}_q^{2}\hat q^{2} + \tilde{\lambda}\hat N\hat q
\end{equation}
Here, we find two effective frequencies: (i) the effective frequency of the vibrational normal mode that is given by 
\begin{equation}
\label{eqn:omegaN_eff}
\tilde{\omega}_{N}^{2} = \omega_{N}^{2} + \Xi + e\lambda^{2}Z\frac{d\braket{\hat \mu}}{dN}
\end{equation}
and (ii) the effective frequency of the photon mode that is given by
\begin{equation}
\tilde{\omega}_{q}^{2} = \omega_{q}^{2} - \lambda\omega_{q}\frac{d\braket{\hat \mu}}{dq}.
\end{equation}
In addition, we have the effective interaction strength that is given by
\begin{equation}
\label{eqn:lamb_eff}
\tilde{\lambda} = -\lambda\omega_{q}\frac{d\braket{\hat \mu}}{dN}
\end{equation}

The resulting eigenvalues are then the upper and lower polaritons with frequencies
\begin{equation}
  \omega_{\pm} =\frac{\tilde{\omega}_{q} + \tilde{\omega}_{N}}{2} \pm \sqrt{\tilde{\lambda}^{2} + \left(\frac{\tilde{\omega}_{q} - \tilde{\omega}_{N}}{2}\right)^{2}}
\end{equation}
We find that the photon frequency $\omega_{q}$ at which resonance
occurs is then not that of the bare phonon mode ($\omega_{N}$), instead resonance
occurs when $\tilde{\omega}_{q} = \tilde{\omega}_{N}$.
The model of vibro-polaritons in Eq.~\ref{eqn:general_model} contains three parameters which have a dependence on the coupling
strength $\lambda$. These are the derivatives of the dipole with respect to
photon displacement $d\braket{\hat\mu}/d q$ and nuclear positions
$d\braket{\hat\mu}/d N$, as well as derivatives of the Coulomb forces on nuclei
expressed as $\Xi$, where derivatives with respect to nuclei positions are in
a basis of uncoupled vibrational normal modes. For an uncoupled system ($\lambda=0$) both $\Xi$ and
$d\braket{\hat\mu}/d q$ are zero. The $\lambda$ dependence of all three of
these parameters is a result of coupling strength and $q$ dependence of the
electronic state. Changes in the electronic state with $\lambda$ and $q$ change
the force terms written as expectation values (i.e. within $\braket{..}$) in
Eqs.~\ref{eqn:F_R} and \ref{eqn:F_q}. This effect is then captured by these
$\lambda$ dependent parameters of the model, written in the basis of vibrational normal modes
of the uncoupled system.

An alternative approach for treating vibro-polaritons from first principles is to
use a model which couples the cavity photon mode to matter vibrations. The parameters, which characterize the matter vibrations, are then obtained from standard first principles
methods~\cite{George2016,kadyan2021,fischer21_groun_state_proper_infrar_spect,hernandez2019}. In
such an approach the modification of the electronic potential energy due to the
cavity is not taken in to account consistently. Such models correspond to neglecting the
coupling strength dependent terms in Eq.~\ref{eqn:general_model}. Setting
$\Xi_{IJ}$ and
$\frac{\partial\braket{\hat{\vec{\mu}}}}{\partial q_{\alpha}}$ equal to zero and
setting $\frac{\partial\braket{\hat{\vec{\mu}}}}{\partial N_{I}}$ equal to its
$\lambda=0$ value recovers such a model which can be constructed without
cavity modification of the electronic potential energy. We will refer to this approximation
as the ``$\mu^{2}$ model'' as it still contains quadratic dipole terms from
Eq.~\ref{eqn:coupling}. If one further neglects this term in the
$N_{I}N_{J}$ coupling that is of order $\lambda^{2}$ one arrives at a
system of bilinearly coupled vibrational and photon oscillators similar to
the Hopfield model~\cite{PhysRev.112.1555}. In this simplified model the single vibration - single photon effective frequencies are simply the bare vibrational and cavity normal modes and any $\lambda$ dependence of $\frac{d\braket{\hat{\vec{\mu}}}}{dN}$ is neglected in the effective coupling strength term.

\section{Results and Discussion}

In this section, we illustrate the developed approach on single and many CO$_2$ molecules, as well as the iron-pentacarbonyl Fe(CO$_5$). We list the numerical details for these calculations in the appendix \ref{sec:num-details}. We start by discussing the case of CO$_2$ molecule(s).

\subsection{Single CO$_2$ in an optical cavity}

\begin{figure}

  \includegraphics[width=\linewidth]{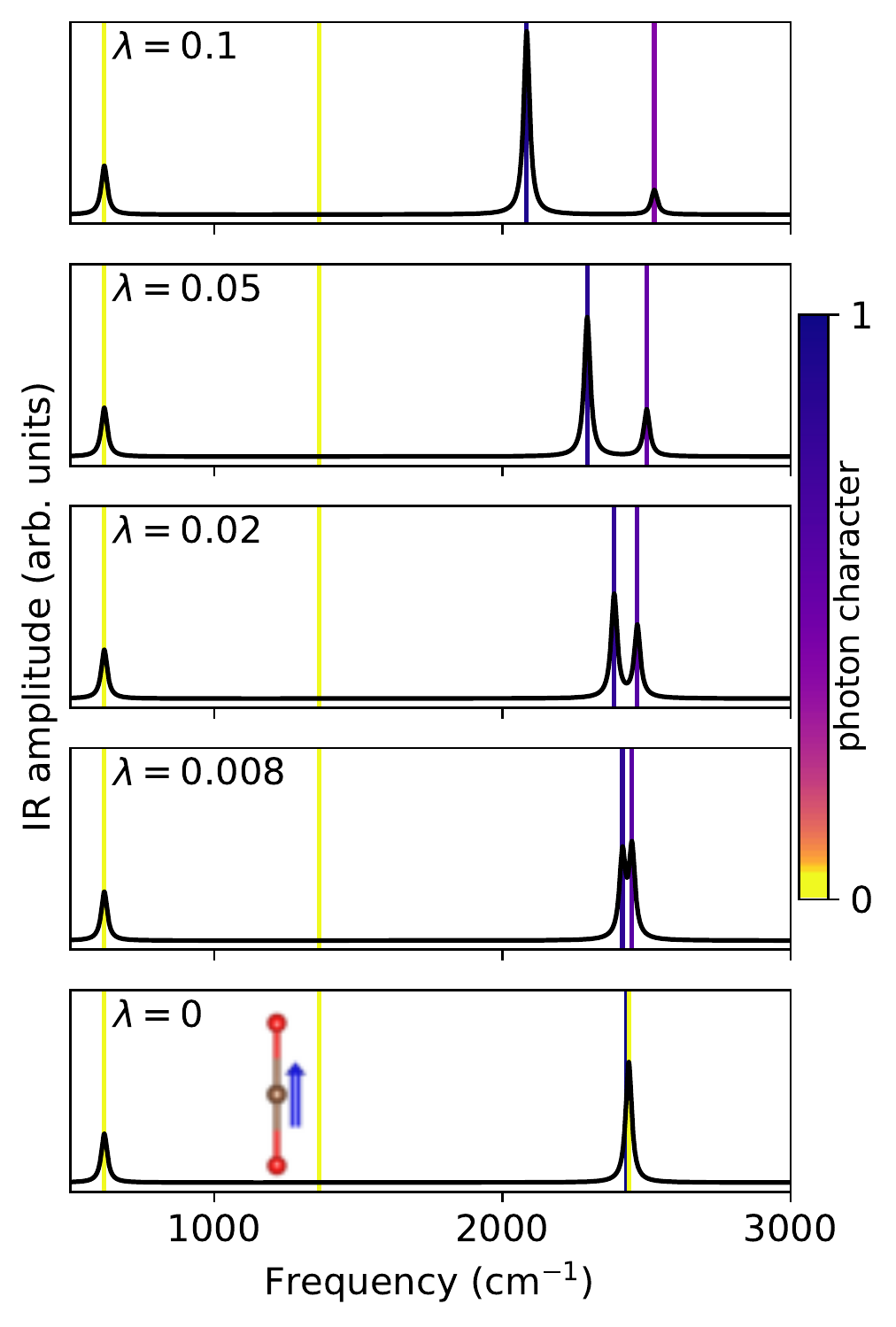}
  \caption{CO$_{2}$ IR spectra for different $\lambda$ values (black curve) for cavity frequency $\omega_\alpha = $ 2430 cm$^{-1}$ and eigenvalues (vertical lines) colored by photonic character. The inset in the $\lambda=0$ plot shows the CO$_{2}$ molecule with the blue arrow indicating the polarization of the photon mode.}
  \label{fig:co2_ir_v_lambda}
\end{figure}

Fig.~\ref{fig:co2_ir_v_lambda} shows the computed vibro-polariton normal mode
frequencies (vertical lines) and Lorentzian broadened infrared
spectra (black curves) at various values of the coupling strength $\lambda$. The
color of the vertical lines corresponds to the absolute value of the photon
component of the corresponding vibro-polariton normal mode eigenvector. In this calculation, one photon
mode is included with frequency $\omega_\alpha=2430$ cm$^{-1}$ chosen to be near
resonance with the 2436 cm$^{-1}$ asymmetric stretching vibration mode of the
uncoupled system. We choose this slight detuning to be consistent with the calculation in Ref.~\cite{flick18_cavit_correl_elect_nuclear_dynam}. The direction of the $\vec{\lambda}_\alpha$ vector, which sets the photon mode polarization direction, was chosen to be aligned with
the oscillating dipole moment along the C-O bonds as indicated by the blue arrow
in the inset of the bottom plot in Fig~\ref{fig:co2_ir_v_lambda}. By increasing the coupling strength from $\lambda = 0$ to $\lambda = 0.1$, we observe the Rabi splitting
of the vibrational mode at 2436 cm$^{-1}$ between the upper and lower vibro-polariton branches. We note that the observed values are in quantitative agreement with the fully time dependent results of Ref.~\cite{flick18_cavit_correl_elect_nuclear_dynam}. As expected, neither the 
non-infrared (IR) active symmetric stretching mode at 1363 cm$^{-1}$ or the degenerate
bending modes at 607cm$^{-1}$ couple to the cavity. The latter of which is only
IR active along directions orthogonal to the cavity polarization. The
eigenvectors of the two polariton modes are linear combinations of the asymmetric
stretching mode and the photon displacement. The lower polariton has a
photon displacement aligned with the vibration mode dipole and a larger photon
component. While for the upper polariton eigenvector the photon displacement
is anti-aligned with the vibration mode dipole and the photon component is
smaller. A coupling strength of $\lambda=0.05$ marks the onset of the strong coupling regime with splitting that is 8.5\% of the uncoupled photon mode. At $\lambda=0.1$ the system is well in to the ultra strong coupling regime with a splitting over 18\% of the uncoupled photon mode.

Notable in the results is the asymmetry in the Rabi splitting, especially in the
strong coupling regime. The lower polariton is
seen to have a more intense IR peak and a larger frequency shift with respect to the frequency of the bare photon mode than the upper polariton. This behavior is despite the finding that the lower
polariton having a smaller matter and larger photon contribution than the upper polariton as can be seen from the peak color. We find however that the IR amplitudes here are
dominated by the change in the electronic contribution to the dipole moment due
to the change in photon displacement $q$, i.e. the term $\frac{\partial \braket{\mu_{\kappa}}}{\partial q_{\alpha}}q_{\alpha}$ in Eq.~\ref{eqn:born-charge}. While the derivative of the dipole moment with respect to $q$ is
smaller in magnitude than the corresponding matter contribution (the Born
effective charges) the photon component of the eigendisplacements can be much
larger than the matter components as the photon components are not reduced by a
factor relating to their mass (from $\tilde{M}$ of Eq.~\ref{eqn:gen_ev_mat} and Eq.~\ref{eqn::gen_ev_mat_compact}). Interestingly, Refs.~\cite{flick2019lr,yang2021quantumelectrodynamical} also show similar behavior for the case of strong coupling to an electronic excitation. In contrast, it is seen in  Ref.~\cite{fischer21_groun_state_proper_infrar_spect} that for the case of a LiH molecule the
lower polariton has larger matter contribution than photon contribution.
However, in that work changes in (electronic) dipole moment due to the cavity
mode displacement are not accounted for so the lower polariton instead ends up
with a smaller peak.

The asymmetry in the frequency splitting for the upper and lower polaritons can
be understood by examining the two mode model presented in
Eqs.~\ref{eqn:two_mode_model}-\ref{eqn:lamb_eff}. The $\lambda$ dependent
parameters $\Xi$, $\frac{d\braket{\mu}}{dN}$, and
$\frac{d\braket{\mu}}{dq}$ enter in a manner which shifts the effective
frequencies of both the vibrational and photon modes. Then even when the cavity
mode is tuned to the frequency of the vibration mode these effective frequencies
differ and thus splitting is not symmetric around the original vibration frequency. The
$\lambda$ dependence of each of these terms is a result of the electronic
response to the cavity potential.

\begin{figure}
  \includegraphics[width=\linewidth]{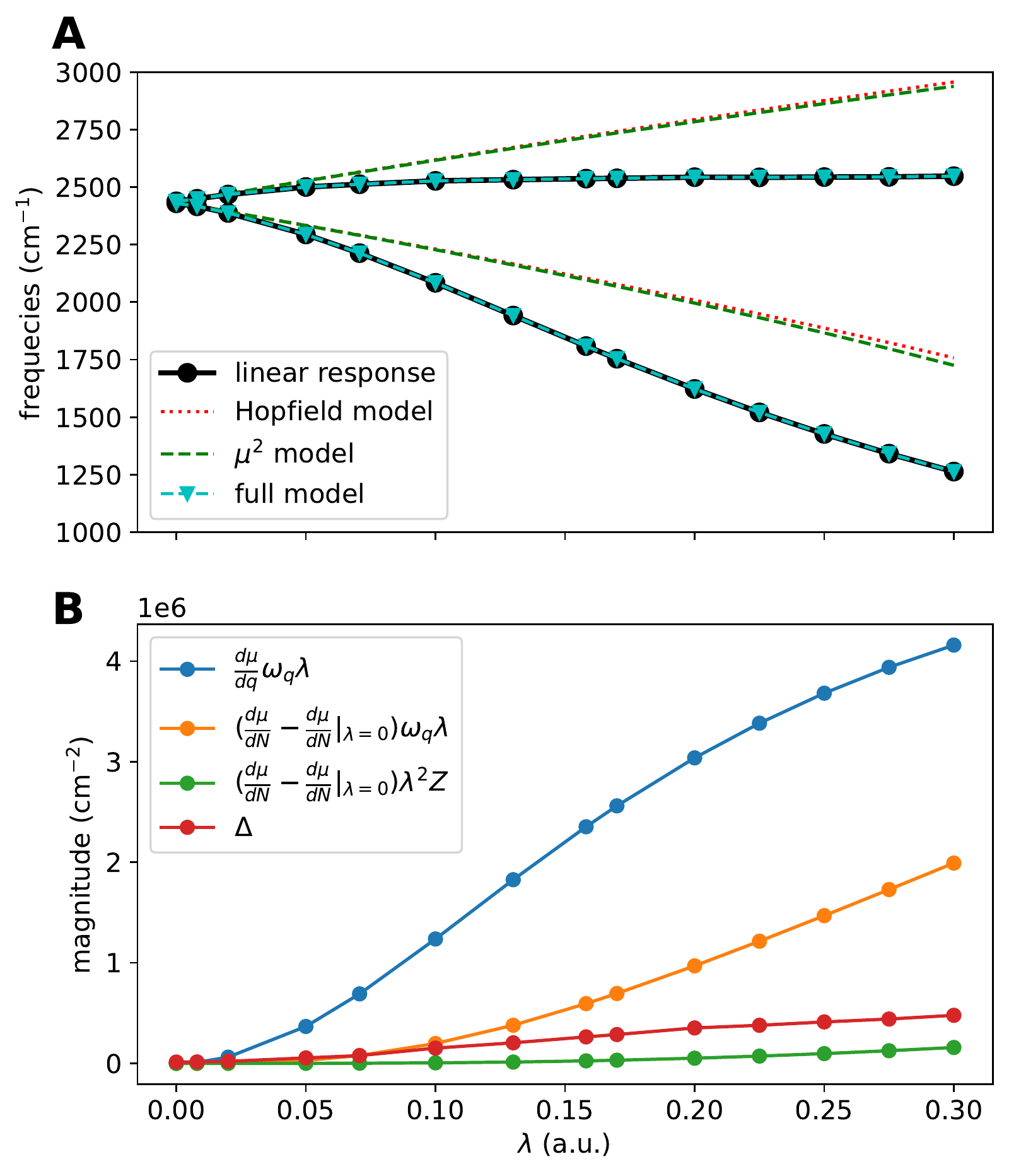}
  \caption{Top (A): CO$_{2}$ mode splitting at various levels of modeling, see
    main text for definitions with cavity frequency $\omega_\alpha = 2430 $
    cm$^{-1}$. Bottom (B): Change of the different model parameters with
    coupling strength $\lambda$.}
  \label{fig:CO2_models}
\end{figure}

In the next step, we compare different effective models to the discussed first principle results, and analyze the individual terms in Eq.~\ref{eqn:two_mode_model} in more detail. Fig.~\ref{fig:CO2_models} A compares the upper and lower polariton frequencies at
different levels of modeling as a function of coupling strength. The results of
the generalized dynamical matrix approach using the first principle theory described in Sec.~\ref{sec:theory}
(shown in black) are seen to be in near perfect agreement with the full two mode
model (shown in blue) of Eq.~\ref{eqn:two_mode_model}. In addition, we compare to two additional approximate models, which show significant differences  from the full model in the ultra strong coupling regime. The first model, shown in green dotted lines, which we term the {$\mu^2$ model}, corresponds to results where all of the $\lambda$ dependence of the model parameters in Eq.~\ref{eqn:two_mode_model} have been neglected so that $\Xi=d\braket{\mu}/dq=0$ and
$d\braket{\mu}/dN$ are taken as the value from the uncoupled case. The second model, shown in the orange dotted line corresponds to a Hopfield type model where in addition to the approximations made for the $\mu$ model also the $\lambda^2$ term from Eq.~\ref{eqn:omegaN_eff} is also set to
zero (equivalent to dropping the $\mu^2$ term in $V_{\mathrm{pt}-\boldsymbol{\mu}}$). For a cavity mode precisely in resonance to a vibration mode the
Hopfield model maintains perfectly symmetric splitting up to the extremely
strong coupling regime. While the inclusion of the $\mu^2$ term does permit some
asymmetry in the splitting it is seen that when parameterized by first
principles results from the $\lambda=0$ limit this asymmetry is relatively
minimal and results do not differ much from the Hopfield model. While some asymmetry is also present due to the small detuning of photon and vibration mode in our calculations, it is only when
coupling dependent model parameters obtained from QEDFT are included that the
more dramatic asymmetric splitting is recovered. Fig.~\ref{fig:CO2_models} B
shows how various terms in the model vary with coupling strength $\lambda$. The
largest $\lambda$ dependent contribution is seen to come from the
$d\braket{\mu}/{dq}$ term. This change in electronic dipole moment due to the
photon displacement shifts the effective cavity mode frequency away from
resonance with the phonon mode.

\subsection{Collective strong-coupling limit with many CO$_2$ molecules}
\subsubsection{First principles results}
\begin{figure}
  \includegraphics[width=\linewidth]{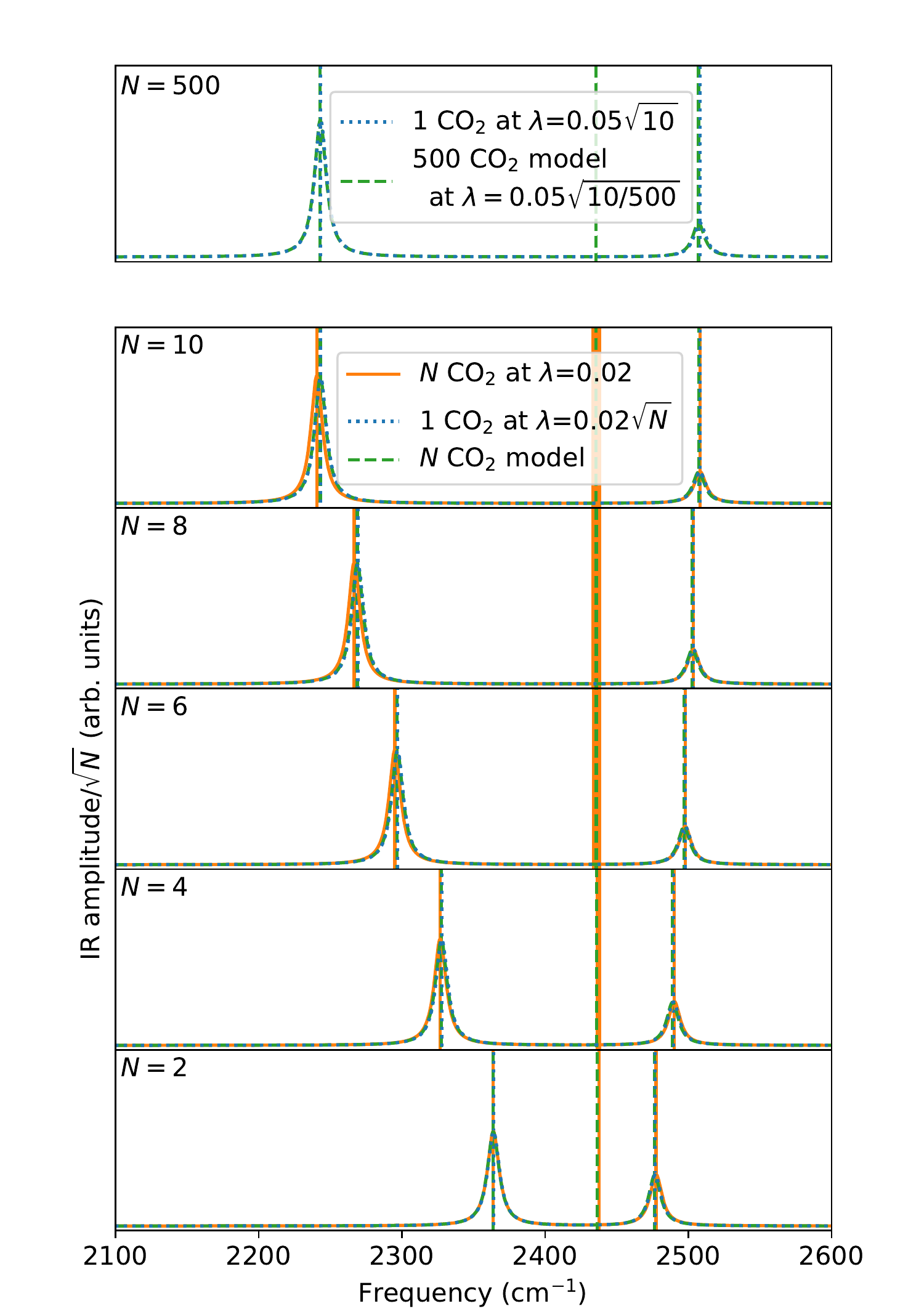}
  \caption{Comparisons between full QEDFT results for $N_\mathrm{mol} $CO$_2$ molecules at
$\lambda=0.05$ (in orange) with results for single CO$_2$ molecules at
$\lambda=0.05\sqrt{N_\mathrm{mol}}$ (in blue). Also shown (in green) are results
for a model of the form in Eq.~\ref{eqn:general_model} for the $N_\mathrm{mol}$
case, but constructed using parameters from QEDFT calculations with only two
CO$_2$ molecules. Vertical lines are used to indicate frequencies of the normal
modes, while curves show broadened IR spectra. To facilitate comparison the IR
amplitudes have been scaled by $N_{\mathrm{mol}}^{1/2}$.}
  \label{fig:multiCO2}
\end{figure}

Rigorous first principles approaches in the treatment of strong light matter
coupling have largely been applied to the problem of a single molecule strongly
coupled to cavity photon modes. However, experimentally strong coupling is
typically achieved via ``collective coupling'' where coupling strength is enhanced
by increasing the number of emitters in the cavity~\cite{sidler20_polar_chemis}.
The increased computational efficiency of the linear response method presented
in Sec.~\ref{sec:theory} enables some aspects of the collective coupling regime
to be accessible within QEDFT. We have simulated chains of CO$_2$ molecules aligned along
their C-O bond directions coupled to a cavity mode polarized along this same
direction. Molecules are chosen to be spaced 20 Bohr apart to simulate the
dilute gas limit. Fig.~\ref{fig:multiCO2} shows comparisons between QEDFT results for a single molecule,
$N_\mathrm{mol}$ molecules, and the results of the many molecule model presented in Sec.~\ref{sec:multimol_model}.
In each of these plots
one can see the lower and upper polaritons similar to those observed in the
single molecule case, but also $N_\mathrm{mol}-1$ dark modes near 2436 cm$^{-1}$
with no IR amplitude. The Rabi splitting and IR spectra in the very strongly
coupled single molecule case and more weakly coupled $N_\mathrm{mol}$ case are
nearly identical with only some differences in the lower polariton frequencies
at very large number of molecules/very strong coupling.

Similar to the case of a single coupled molecule the lower (upper) polariton
eigendisplacements consist of the original asymmetric stretching mode aligned
(antialigned) with the photon displacement. However, now in the multi-mode case
the collective upper and lower polaritons consist of every molecule experiencing
this asymmetric stretching in phase. The multi-molecule setup also results in a number of
dark modes which correspond to combinations of the original asymmetric
stretching modes on each molecule, but in such a way that the overall dipole
moment when freezing in one of these collective dark modes is zero.

\subsubsection{Modelling larger numbers of molecules}
\label{sec:multimol_model}

The similarity between the results of a single strongly coupled molecule with
multiple, more weakly coupled molecules suggests that within the level of theory
applied in this work the microscopic description of one or two molecules can
capture the relevant physics for many molecules coupled to the cavity in the
dilute limit. To this end we construct a model of the form presented in
Eq.~\ref{eqn:general_model} with $N_{\mathrm{mol}}$ CO$_{2}$ molecules coupled
to the same cavity mode as in previous sections at a coupling strength of
$\lambda^{(N_{\mathrm{mol}})}$. Nearly all parameters in this model can be
obtained from first principles calculations of a single molecule with coupling
strength $\lambda^{(1)}=\sqrt{N_{\mathrm{mol}}}\lambda^{(N_{\mathrm{mol}})}$
except for certain elements of the $\Xi$ matrix which we obtain from first
principles calculations for two molecules with coupling strength
$\lambda^{(2)}=\sqrt{N_{\mathrm{mol}}/2}\lambda^{(N_{\mathrm{mol}})}$\footnote{All
  parameters can be obtained from the two molecule calculation, but for clarity
  we present the parameters which can be obtained from a single molecule
  calculation as coming from such a calculation}. The $N_{\mathrm{mol}}$ model
consists of the same photon modes as single molecule case so
$\omega^{(N_{\mathrm{mol}})}_{\alpha} = \omega^{(1)}_{\alpha}$ and
$N_{\mathrm{mol}}$ copies of the vibration modes from a single uncoupled
molecule.
To simplify the notation for mapping model parameters of the $N_{\mathrm{mol}}$
system to the parameters of corresponding one or two model parameters we have
introduced the superscript indicating the number of molecules in the model a
particular parameter corresponds to. Since we will be including copies of the
original, single molecule, vibrational modes as our starting basis it is convenient to write our
nuclear degrees of freedom with two indices; a molecular index $\mathbb{M}$ and
vibrational mode index $I$ which corresponds to a normal mode of the
uncoupled single molecule system. Together the pair of indices $(\mathbb{M}I)$
corresponds to an atomic displacement on molecule $\mathbb{M}$ according to the
eigendisplacement of the single molecule vibrational mode given by
$\eta^{(\uline{\vec{R}})}_{I}$.
 So for $N_{\mathrm{ions}}$ ions
in each molecule in three dimensions
$\omega^{(N_{\mathrm{mol}})}_{(\mathbb{M}I)} = \omega^{(1)}_{I}$,
$Z^{(N_{\mathrm{mol}})}_{(\mathbb{M}I)} = Z^{(1)}_{I}$,
and
${(\frac{d\braket{\mu}}{dN_{(\mathbb{M}I)}})}^{(N_{\mathrm{mol}})} = {(\frac{d\braket{\mu}}{dN_{I}})}^{(1)}$.
Within the dipole approximation a change in $q_{\alpha}$ will result in a change
in dipole moment for all molecules in the system so the susceptibility must be
scaled for the model as
${(\frac{d\braket{\mu}}{dq_{\alpha}})}^{(N_{\mathrm{mol}})} = \sqrt{N_{\mathrm{mol}}}{(\frac{d\braket{\mu}}{dq_{\alpha}})}^{(1)}$.
For the choice of basis consistent with the above definitions $\Xi$ has a
block structure where on diagonal blocks correspond to coupling between vibration modes on
the same molecule and off diagonal blocks correspond to coupling between
vibration modes of different molecules. While the on diagonal blocks can be
obtained via ab-initio calculations on a single molecule the latter requires ab
initio treatment of two molecules. The details of this construction are presented in Appendix~\ref{appendix:delta}.
Since the molecules are sufficiently separated and since the long range $\mu^2$
term is in practice handled with the mean field approximation of
Eq.~\ref{eqn:vmf} the impact of any one molecule on another is nearly
independent of their distance. And the impact of two molecules on a third is
equivalent to a single molecule contributing the same change in dipole moment.
So to harmonic order the case of two molecules captures nearly all relevant
interactions to describe $N_\mathrm{mol}$ molecules within the level of theory
used in this work.

It is seen that within the dipole approximation there is
almost no difference in the IR spectrum between a single molecule strongly coupled and a collection
of molecules more weakly coupled aside from the appearance of dark modes.
However, the coupling used in Eq.~\ref{eqn:coupling} when applied to the many
molecule case assumes equal coupling to all molecules in the system as there is
no spatial dependence of $\vec{\lambda}_\alpha$. A more realistic simulation of
collective coupling would facilitate better understanding of the similarities
and differences between local and collective strong coupling and will be the
subject of subsequent work.

\subsection{Fe(CO$_5$) in multiple photon mode setup}
 \label{ssec:FeCO5}
In the previous section a single cavity mode was coupled to numerous degenerate
vibration modes each on different molecules. In this section we investigate a
cavity coupled to multiple degenerate and non-degenerate vibration modes of a
single iron-pentacarbonyl molecule. Experimental data of a similar system setup has been published in Ref.~\cite{George2016}.

\begin{figure}
  \includegraphics[width=\linewidth]{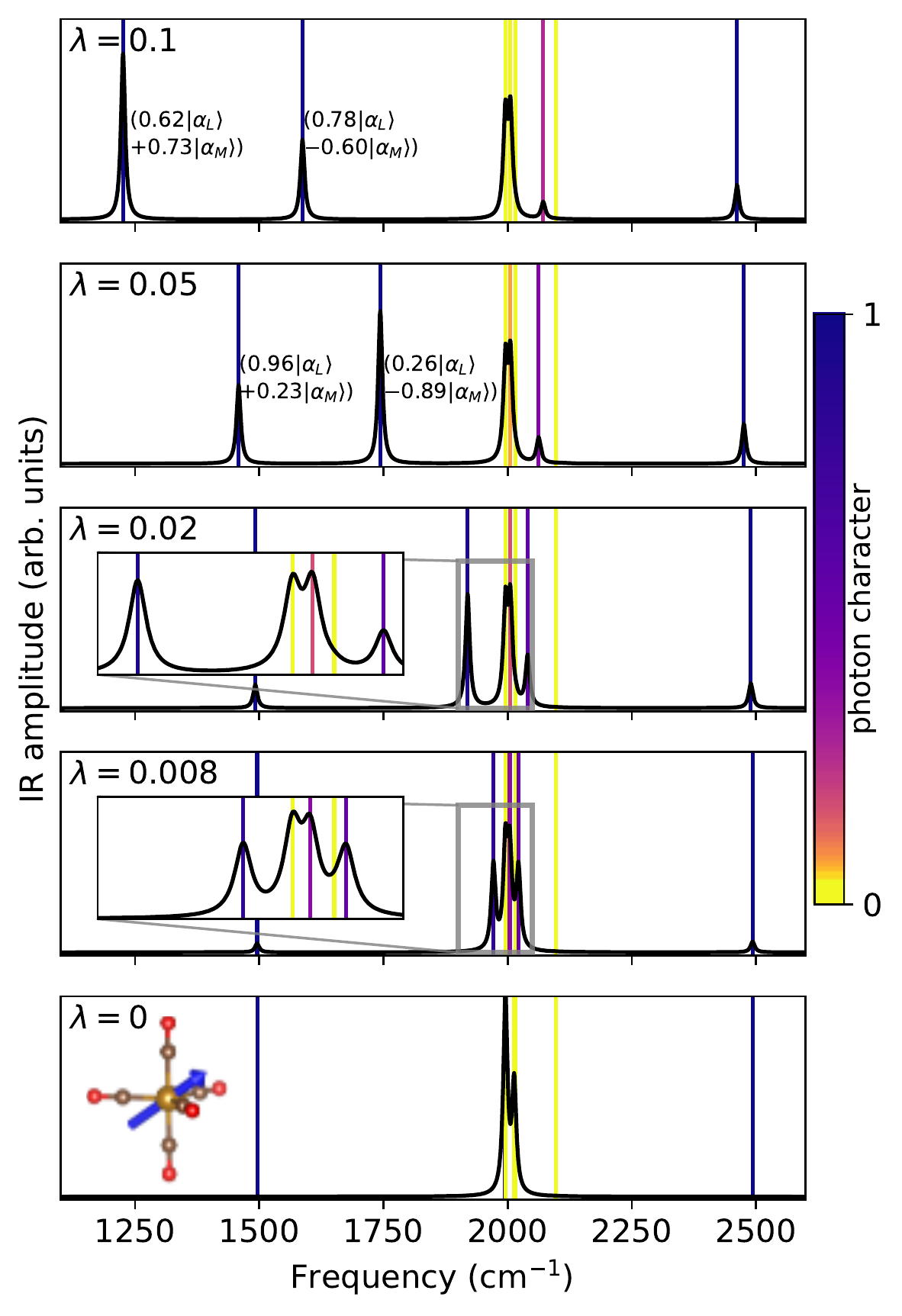}
  \caption{Fe(CO)$_{5}$ IR spectra for different $\lambda$ values (black curve)
    and eigenvalues (vertical lines) colored by photon character. Here three
    photon modes are present, corresponding to 3rd, 4th, and 5th harmonics of
    the optical cavity.    The $\lambda$ value indicated in the plot is
    that of the 4th harmonic (at 1995 cm$^{-1}$), the other two harmonics have
    been set to have $1/3$rd the coupling strength of the 4th harmonic. The
    inset of the $\lambda=0$ plot shows the Fe(CO)$_{5}$ molecule, with the blue
    vector indicating the direction of the photon mode polarization.
    Annotations give the values of the projections of the vibro-polariton eigenvectors on to the 3rd and 4th harmonic uncoupled photon states.}
  \label{fig:feco_ir_v_lambda}

\end{figure}

Our system is studied with a cavity mode in
resonance with several IR active vibrations as well as with two additional
photon modes to simulate additional harmonics of the
cavity. As shown in the inset of the bottom panel of
Fig.~\ref{fig:feco_ir_v_lambda} the coupled cavity polarization is set to be
along an axis 45 degrees from the axis of the 3 fold rotational symmetry. This
setup leads to a coupling to both the vibrational mode at 2013 cm$^{-1}$ which involves
polar distortions along the 3-fold axis and the two degenerate vibrational modes at 1995
cm$^{-1}$ which involve distortions perpendicular to the 3-fold axis. A cavity
mode at 1995 cm$^{-1}$ is set to couple most strongly while additional
``harmonics'' at frequency ratios of 3/4 (1496 cm$^{-1}$) and 5/4 (2494
cm$^{-1}$) are set to have a coupling strength 0.3 times that of the
central mode. 
There are also two non-IR active vibrational modes nearby in energy at 2016 cm$^{-1}$ and
2097 cm$^{-1}$ which do not couple to any cavity modes.
Fig.~\ref{fig:feco_ir_v_lambda} depicts the normal modes of the system as
vertical lines colored by photon character as well as the Lorentzian broadened
IR spectra at several coupling strength magnitudes. 

At coupling strengths with $|\lambda|<=0.02$ the two outer cavity modes at 1496
cm$^{-1}$ and 2494 cm$^{-1}$ are approximately uncoupled from the vibrational
modes of the system and the central photon mode at 1995 cm$^{-1}$. The IR amplitudes for
the outer modes in the regime are dominated by the effect the cavity mode has on
the electronic system (through the $d{\mu}/{dq}$ term). The central cavity photon mode
interacts the three IR active vibrational modes nearby in energy; the polar along the 3-fold
axis (z) mode at 2013 cm$^{-1}$ and the two degenerate polar modes within the
plane perpendicular to the 3-fold axis (xy) modes at 1995 cm$^{-1}$. The result
of this cavity induced coupling is four nondegenerate modes; a dark state which
is a linear combination of the two xy modes, and three polaritons which are
linear combinations of the cavity photon mode, xy modes, and the z mode. The
dark state is still IR active, but not along the cavity mode polarization
direction. Similar to the case with CO$_{2}$ as coupling strength is increased
the frequencies of the upper (lower) most polariton continue to grow larger
(smaller) respectively while the photon character of the polariton mode decreases
(increases). The middle polariton rapidly converges to a frequency of
2006 cm$^{-1}$ and as coupling strength increases the photon character of this
mode decreases until there is no photon character and the mode is
made up of a linear combination of the polar z and xy vibrations. The cavity has
induced a coupling between these polar vibration modes changing the eigenstate even in a
regime where this eigenstate has no photon character.

At extremely strong coupling strengths with $|\lambda|>= 0.05$ the outer cavity
mode harmonics begin to interact with other modes of the system. In the top two
panels of Fig.~\ref{fig:feco_ir_v_lambda} it can be seen that even the lower
frequency IR active modes below 700 cm$^{-1}$ begin to pick up some small photon
character. Furthermore, as coupling strength increases to this very strong
regime the lower polariton has begins to mix with this lower frequency cavity
mode. The two normal modes between 1200 and 1600 cm$^{-1}$ become a linear
combination of vibrations and both the lowest harmonic cavity photon as well as
the central cavity photon. At $|\lambda|=0.1$ we observe that this effective
photon-photon interaction has grown so strong that the photon mode components of
these two modes essentially swap so that the eigenvector of the mode at 1226
cm$^{-1}$ has a larger component coming from the cavity photon mode at 1995
cm$^{-1}$ and the mode at 1587 cm$^{-1}$ has a larger component from the cavity
mode at 1496 cm$^{-1}$.

\section{Summary and Conclusion}

In this work, we have introduced a first principles framework to calculate the vibro-polaritonic
normal modes of systems when light and matter are strongly coupled. 
Employing the cavity-Born Oppenheimer approximation to separate
electronic from nuclear and photonic degrees of freedom and constructing
dynamical matrices that include the photonic degree of freedom 
enables us to characterize these vibro-polariton states. 
Our approach is based
on QEDFT, which makes it applicable to large system sizes while including effects of the cavity on electronic states. We demonstrate the
framework on calculations for single and many CO$_2$ molecules and iron
pentacarbonyl Fe(CO$_5$). In addition, we derive and compare to a first-principles
based model, that allows for the extrapolation of first principle calculations of few
molecules to the collective strong coupling limit of molecular ensembles. 

Our work opens many different avenues to explore.
The techniques used here can be extended to other properties related to the system normal modes such as the low frequency Raman spectra.
The vibro-polaritonic normal modes computed using the methods developed could be used as an efficient
basis for exploring anharmonic couplings including interactions between polaritonic excitations~\cite{juraschek2019cavity}.
The collective setup employed
in this work assumes the same coupling strength for all molecules. A more realistic description where
different molecular positions imply different coupling strength due to the
profile of the cavity mode could provide insight in to potential differences between
the collective coupling limit and small numbers of very strong coupled molecules. 
Such techniques can also be used to explore other related questions such as the engineering of
 strong coupling on single atoms~\cite{schutz2020} and local modifications
of impurities due to collective coupling~\cite{siedler2021}.
We have utilized the cavity Born-Oppenheimer approximation and treated the
electronic portion of the two body operator $\hat{\mu}^2$ via a mean field potential.
More sophisticated treatment of exchange-correlation effects both of electron-photon
interactions and how the presence of the cavity can modify electron-electron interactions are of interest. Such more advanced treatments will be especially important when energy surfaces are sufficiently close together and the validity of the CBOA should be carefully tested.
Utilizing the methods developed in this work, potentially along with these extensions, experimentally relevant molecules can
be studied to gain new insights on cavity modification of chemical reactivity.
Also of interest is the extension of QEDFT approaches, including the linear response technique presented here, to solid state systems treated with periodic boundary
conditions to study the effects of optical cavities
on phonons and phonon-polaritons~\cite{latini2021ferroelectric}.

\section{Acknowledgements} All calculations were performed using the computational facilities of the Flatiron Institute. The Flatiron Institute is a division of the Simons Foundation.

\section{Appendix}

\appendix

\section{ Numerical details}
\label{sec:num-details}

We have implemented the presented method into the real-space and pseudopotential
time-dependent density-functional theory code Octopus
~\cite{octopus1,octopus2, octopus3} and will be made publicly available in a future release.
Calculations were performed with the PBE exchange-correlation functional \cite{PhysRevLett.77.3865}
using optimized norm-conserving Vanderbilt pseudopotentials \cite{VANSETTEN201839, PhysRevB.88.085117} 
on a real space grid with spacing 0.1 \AA \, and simulation box edges with  at least 4 \AA\, distance from the center of each ion.
 To describe the derivatives in
Eqs.~\ref{eqn:CRR}-\ref{eqn:CqR}, we use the finite-difference procedure, i.e.
calculate total energy differences for different nuclear and photonic
displacements, respectively.

\section{ Mean field electronic potential}
\label{appendix:vmf}
In practice to obtain the necessary force components in the matrix equation of Eqs.~\ref{eqn:CRR}-\ref{eqn:CqR}, one must solve the electronic system defined by Eq.~\ref{eqn:cboa-ham}. In this work, we use a mean-field approximation to describe the $R^2$ term~\cite{flick18_cavit_correl_elect_nuclear_dynam,ruggenthaler14_quant_elect_densit_funct_theor} given explicitly by the following equation
\begin{equation}
\label{eqn:vmf}
  \begin{split}
  \hat{V}_{\mathrm{pt}-\vec{\mu}}^{(\mathrm{MF})}(\vec{r},\vec{\underline{R}}, {\underline{q}}) &=
 \sum_{\alpha=1}\frac{1}{2} \omega_\alpha^2 q_\alpha^2  + \frac{1}{2}\lambda_\alpha^2\left(\sum_I e Z_I \vec{R}_I\right)^2 \\
  &- \lambda_\alpha \omega_\alpha q_\alpha \left(\sum_I e Z_I \vec{R}_I\right) \\
 & -e\vec{r}\biggl[\lambda_\alpha^2 \left(\sum_I e Z_I \vec{R}_I - \int d^3r e \vec{r} \rho(r)\right) \\
    & - \lambda_\alpha \omega_\alpha q_\alpha \biggr]\\
  &+ \frac{1}{2}\lambda_\alpha^2\left(\int d^3r e \vec{r} \rho(\vec{r})\right)^2
  \end{split}
\end{equation}

\section{Model derivation}
\label{appendix:model_details}

The first force contribution on the left side of Eq.~\ref{eqn:F_R} becomes
\begin{equation}
  \label{eqn:Delta}
  \begin{split}
  \braket{\frac{d}{dN_{I}} (\hat{V}_{e-\mathrm{nuc}} + \hat{V}_{\mathrm{nuc}-\mathrm{nuc}})}
  &\approx \sum_{J}(\omega_{I}^{2}\delta_{I, J} + \Xi_{I, J})N_{J}\\
  &+ \sum_{\alpha}\Theta_{\alpha, I}q_{\alpha}
  \end{split}
\end{equation}
where $\omega_I$ is the frequency of the uncoupled vibration mode and we have expressed mixed second derivatives of this force contribution in
terms of new matrices $\Theta$ and $\Xi$, defined by
\begin{equation}
  \Xi_{I,J} =
  \frac{d}{dN_{I}}\braket{\frac{d}{dN_{J}} (\hat{V}_{e-\mathrm{nuc}} + \hat{V}_{\mathrm{nuc}-\mathrm{nuc}})} - \omega_{I}^{2}\delta_{IJ}
\end{equation}
and 
\begin{equation}
  \Theta_{\alpha, I} =
  \frac{d}{d q_{\alpha}}\braket{\frac{d}{dN_{I}} (\hat{V}_{e-\mathrm{nuc}} + \hat{V}_{\mathrm{nuc}-\mathrm{nuc}})}.
\end{equation}

The $\Xi$ matrix can be constructed from the force contribution on the left
side of Eq.~\ref{eqn:F_R} by changing the basis to that of the uncoupled normal
vibration modes as follows.
We can first define $\mathbf{B}$
\begin{equation}
B_{I \kappa, J \kappa'} = \frac{1}{\sqrt{m_{I}m_{J}}}\frac{d}{d R_{I \kappa}}\braket{\frac{d}{d R_{J \kappa'}}(V-V_{pt})}
\end{equation}

which is similar to the dynamical matrix, but neglecting explicit contributions from
the coupling term (though they do impact the electronic state and thus the
expectation value). We then obtain $\Xi$ by transforming $\mathbf{B}$ into
the basis of dynamical matrix eigenvectors (for the uncoupled system) and
subtracting the bare phonon frequencies.
\begin{equation}
 \Xi_{I,J} = \sum_{ij}U^{\mathrm{T}}_{I i}  B_{ij}  U_{j J} - \delta_{IJ}\omega_{I}^{2}
\end{equation}
Note in the above sum $i,j$ each run over both atom and direction indices (eg. with indexing starting from zero $j=3J+\kappa'$).
Then from Eq.~\ref{eqn:F_R}  we can obtain
\begin{equation}
  F_{N_{I}} = \braket{\frac{d}{dN_{I}} (\hat{V}_{e-\mathrm{nuc}} + \hat{V}_{\mathrm{nuc}-\mathrm{nuc}})}
    + e Z_{i}\sum_{\alpha=1}\vec{\lambda}_{\alpha}\biggl(\omega_{\alpha}\hat{q}_{\alpha}  -\vec{\lambda}_{\alpha}\cdot  \braket{\hat{\vec{\mu}}}\biggl)
\end{equation}
where the first term on the right is given in  \ref{eqn:Delta}.

We can then use Eqs.~\ref{eqn:F_R} and \ref{eqn:F_q},  to obtain the model from

\begin{align}
  \label{eqn:E_Rq_force}
    E(\vec {\underline{R}}, \underline{q}) &= E_0 +
    \sum_{I J}  \frac{1}{2}
     (-\frac{dF_{N_{I}}}{dN_{J}})
     N_{I} N_{J}  \nonumber\\
    &+ \sum_{\alpha,\alpha'}  \frac{1}{2}
(     -\frac{dF_{q_{\alpha}}}{dq_{\alpha'}})
    \Delta q_\alpha \Delta q_\alpha' \nonumber\\
    &+ \sum_{\alpha,I}
(     -\frac{dF_{q_{\alpha}}}{dN_{I}})
      N_{I} \Delta q_\alpha.
\end{align}

It may seem notable that Eq.~\ref{eqn:general_model} does not contain the
$\Theta$ term present in Eq.~\ref{eqn:Delta}. This is possible due to the relation
\begin{equation}
  \label{eqn:maxwell_reln}
\Theta^{(\uline{\lambda})}_{\alpha I} - \vec{Z}_{I}\cdot(\lambda_{\alpha}\omega_{\alpha} +
\lambda_{\alpha} (\frac{d\braket{\mu}}{dq_{\alpha}}\cdot\lambda_{\alpha}))
= - \omega_{\alpha}\lambda_{\alpha}\cdot\frac{d\braket{\mu}}{dN_{I}}
\end{equation}
which can be seen by setting
$\frac{d F_{R_{I\kappa}}}{d q_{\alpha}} = \frac{d F_{q_{\alpha}}}{d R_{I\kappa}} = - \frac{\partial^{2}E}{\partial q_{\alpha} \partial R_{I\kappa}}$.

\section{Multi molecule delta matrix}
\label{appendix:delta}
To construct $\Xi^{N_{\mathrm{mol}}}$ we make use
of the corresponding matrix from the single molecule case $\Xi^{(1)}$, though
we also need components of this matrix which mix vibration modes of different
molecules. One can utilize $\Xi^{(2)}$, which is constructed using the normal
modes of the two molecule system rotated to a basis of single
molecule normal modes using the dynamical matrix eigenvalues of the uncoupled
single molecule $U^{(1)}$:
\begin{equation}
  \Xi^{(2*)} = {(I \bigotimes U^{(1)})}^{T}\Xi^{(2)}(I \bigotimes U^{(1)}),
\end{equation}
where $I$ is a $2\times2$ identity matrix and the above Kronecker
products promote $U^{(1)}$ to a block diagonal matrix of the same
dimensionality as $\Xi^{(2)}$. So long as the molecules are sufficiently
spatially separated $\Xi^{(2*)}$ will have diagonal blocks identical to
$\Xi^{(1)}$ corresponding to effects arising on a single molecule and some in
general nonzero, but symmetric, off diagonal blocks corresponding to
interactions between molecules. We can adopt the molecular index labeling used
for the $N_{\mathrm{mol}}$ case to refer to these blocks where for this two
molecule case the molecular index only runs over ${0,1}$. Then the model
$\Xi$ matrix for $N_{\mathrm{mol}}$ molecules is given by
$\Xi^{(1)}/N_{\mathrm{mol}}$ on the block diagonal off diagonal blocks and
copies of the off diagonal part of $2\Xi^{(2*)}/N_{\mathrm{mol}}$ on all off
diagonal blocks
\begin{equation}
  \Xi^{(N_{\mathrm{mol}})}_{(mI),(m'J)} =
  \begin{cases}
    \frac{1}{N_{\mathrm{mol}}}\Xi^{(1)}_{I,J} & \text{for } m=m'\\
    \frac{2}{N_{\mathrm{mol}}}\Xi^{(2*)}_{(0I),(1J)} & \text{for } m\neq m'
  \end{cases}.
\end{equation}

\bibliography{references.bib}{}
\end{document}